# A Relation Spectrum inheriting Taylor series：muscle synergy and coupling for hand

Gang Liu, Jing Wang

**Abstract**

There are two famous function decomposition methods in math: Taylor Series and Fourier Series. Fourier series developed into Fourier spectrum, which was applied to signal decomposition\analysis. However, because the Taylor series whose function without a definite functional expression cannot be solved, Taylor Series has rarely been used in engineering. Here, we developed Taylor series by our Dendrite Net, constructed a relation spectrum, and applied it to model or system decomposition\analysis. *Specific engineering*: the knowledge of the intuitive link between muscle activity and the finger movement is vital for the design of commercial prosthetic hands that do not need user pre-training. However, this link has yet to be understood due to the complexity of human hand. In this study, the relation spectrum was applied to analyze the muscle-finger system. One single muscle actuates multiple fingers, or multiple muscles actuate one single finger simultaneously. Thus, the research was in muscle synergy and muscle coupling for hand. This paper has two main contributions. (1) The findings of hand contribute to designing prosthetic hands. (2) The relation spectrum makes the online model human-readable, which unifies online performance and offline results.

Code (novel tool for most fields) is available at https://github.com/liugang1234567/Gang-neuron.

**Taylor Series, relation spectrum, Dendrite Net, prosthetic hands, electromyography**

## 1  Introduction

Myoelectric prosthetic hands, where amputees control prosthetic hand by voluntarily contracting their residual muscles, are attracting considerable critical attention [1]. To our knowledge, they are classified into four types according to decoding ways of electromyography (EMG). *(1)* Type 1 is that two electrodes are attached to the residual muscles, and then the corresponding joint movement is actuated proportionally to the EMG amplitude [2]. *(2)* Further, the function that switches the active joint by a co-contraction of both muscle groups or other heuristics is added to Type 1, which produces Type 2. [3]. *(3)* In order to control more degrees-of-freedoms (DOFs), intensive researches focus on motion classification that assigns EMG features to a discrete set of motions [4, 5]. *(4)* Recently, an approach that maps muscle activation to force or motion by training a complex "black-box" neural network (NN) has been investigated for simultaneous and proportional myoelectric control [1, 6]. Without a doubt, the first type is the most convenient and intuitive. Its intended function corresponds to the physiologically appropriate muscles. Today, most of the commercial devices use this method. However, these devices usually use a two-recording-channel system to control a single DOF because it is unclear about the intuitive link between muscle activity and the finger movement.

• *Gang Liu is with Institute of Robotics and Intelligent Systems, Xi'an Jiaotong University, Shaanxi, P.R.China. [Gang Liu presented relation spectrum.]*

• *Jing Wang is with Institute of Robotics and Intelligent Systems, Xi'an Jiaotong University, Shaanxi, P.R.China.[Jing Wang offered guidance.]*

*How do multiple muscles actuate one single finger?* To solve this problem, we need a white-box algorithm. In 2009, Jiang et al. presented a DOF-wise nonnegative matrix factorization (NMF) algorithm to extract the wrist's neural control information from EMG [7]. Nevertheless, until now, it has not been used for the finger movement, probably because the hand is much more complicated than the wrist. Besides, it is worth emphasizing that although all kinds of machine learning models were used for proportional no-linear myoelectric control, they do not show the intuitive link due to the "black-box" nature [1, 8].

*How does one single muscle actuate multiple fingers?* An early study proves that mechanical coupling and muscle coupling limit finger independence [9]. The mechanical coupling and muscle coupling are useful for the design of myoelectric prosthetic hands [10]. However, the muscle coupling of the finger has not been systematically investigated.

This paper showed the relation spectrum in engineering based on Dendrite Net (DD), the first white-box machine learning algorithm [11]. The Dendrite Net could be found in the previous paper from the same authors [11]. This paper is the first application of DD except for the original algorithm article, and the relation spectrum in engineering is proposed for the first time. Additionally, we gave a proof-of-principle of similarity between the relation spectrum and Taylor series in this paper.

From the Fourier-like perspective [12], Dendrite Net is akin to Fourier transform; the relation spectrum is similar to Fourier spectrum. However, the Fourier spectrum is the decomposition of the signal, and the relation spectrum is the decomposition of the model or system. For the specific engineering problem, this study aims to explore the intuitive link between muscle activity and finger movement. The framework of this engineering problem is shown in Figure 1.

## 2 Materials and Methods

This section describes the relation spectrum in engineering based on Dendrite Net by solving the specific engineering problem of the muscle-finger system.

### 2.1 Subjects

The data analyzed in this paper were from the *scientific data* that includes intramuscular electromyography (iEMG) data related to isometric hand muscle contractions of 14 subjects [13]. These subjects were divided into two protocols: the first focused on the muscles available within a short residual forearm (SRL); the second focused on fingers and thumb muscles (LRL).

The twelve subjects had six pairs of fine-wires inserted regardless of the protocol. However, two other subjects were recorded with nine electrodes whose positions met both SRL and LRL. Thus, the two subjects provided four groups of data and were regarded as four subjects. Thus, we got 8 SRL subject data and 8 LRL subject data. According to the preliminary test and the introduction in [13], we selected the LRL subject that focused on fingers and thumb muscles and named Subject 1-8. The LRL protocol targeted the following muscles: flexor digitorum profundus (FDP), extensor digitorum communis (EDC), abductor pollicis longus (APL), fexor pollicis longus (FPL) - responsible for thumb flexion, extensor pollicis longus (EPL) - responsible for thumb extension, and extensor indicis proprius (EIP) - responsible for index finger (D2) extension [13].

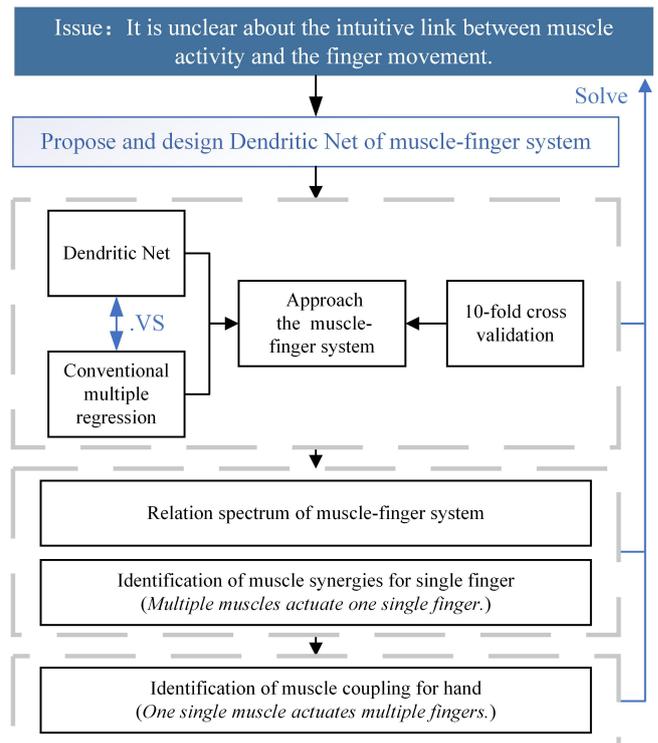

**Figure 1**  Framework of this specific engineering problem.

### 2.2 Acquisition setup and protocol

*1) Acquisition Setup*

Multiple sensors were used to record hand forces and corresponding muscular activity during the experiments. Hand forces were measured using a custom-made force measurement device [13]. The intramuscular EMG signals were recorded using the Quattrocento (OT Bioelettronica, Torino, Italia) biomedical amplifier system. All iEMG signals were sampled with a 16-bit amplitude resolution at 10240 Hz. A hardware high-pass filter at 10 Hz and a low-pass filter at 4400 Hz were used during recordings. The intramuscular electrodes were paired fine-wire electrodes

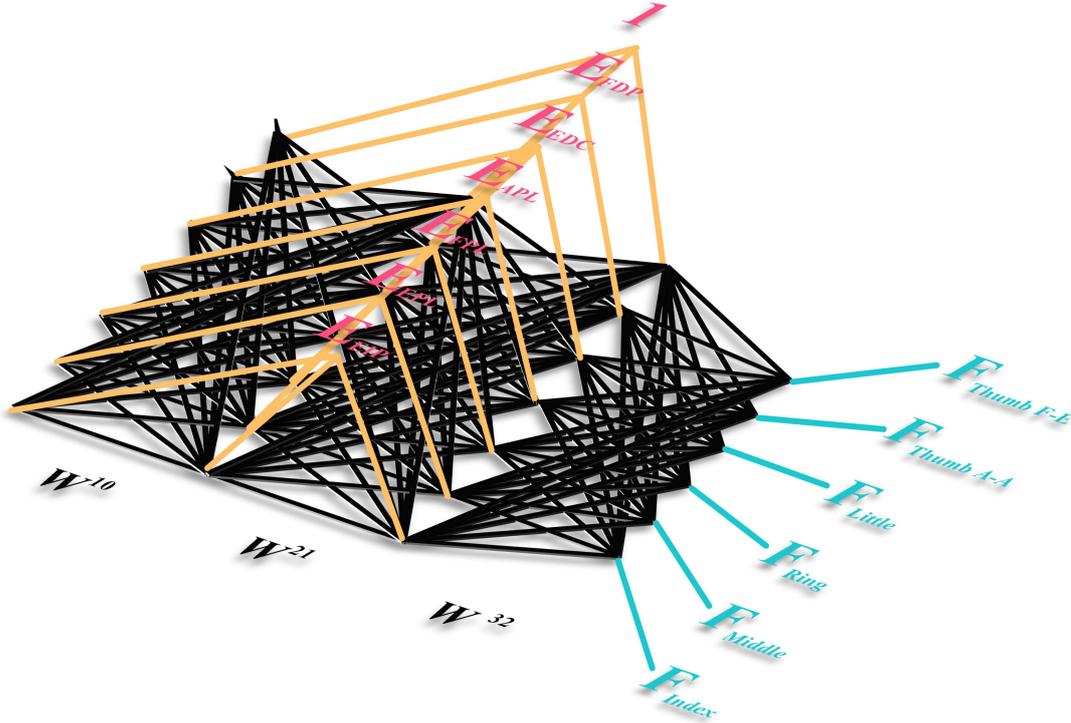

**Figure 2** Dendritic Net architecture of muscle-finger system. $E_{FDP}$: iEMG RMS of FDP; $E_{EDC}$: iEMG RMS of EDC; $E_{APL}$: iEMG RMS of APL; $E_{FPL}$: iEMG RMS of FPL; $E_{FPL}$: iEMG RMS of EPL; $E_{EIP}$: iEMG RMS of EIP.

from Chalgren, Gilroy, USA.

The positionings of the fine-wire electrodes were performed by an MD specialist in clinical neurophysiology using the guidelines from *Anatomical guide for the electromyographer: the limbs and trunk* [13, 14].

2) *Acquisition Protocol* [13]

The subject sat a chair comfortably and was instructed to place the hand in the force measurement device. The whole measurement protocol was controlled and guided automatically by the custom-made software developed in LabVIEW. The subject was asked to produce force/torque that matched the cue presented on the screen. Sinusoidal waveforms were provided as visual cues. The rationale behind the sinusoidal tracking task was to generate a gradual force increase and provide iEMG and force data that described the muscle-finger system. Concretely, the repetition frequency was set to 0.1 Hz to enable a gradual and controllable force increase.

## 2.3 Data Processing

1) *Pre-Processing*

The following steps were executed for all subjects. In order to obtain the iEMG data without 50 Hz noise and its harmonics from, iEMG Data were filtered with a 10 Hz high pass filter, a 450 Hz low pass filter, and a notch filter at 50 Hz [5]. The Root Mean Square (RMS) is one of the most common EMG signal features that represents the signal envelope. In this study, as the same as the literature [13], the RMS was calculated using 250 ms wide window (Matlab command: *square*, *smooth*). Then, we obtained the iEMG data and force data of single DoF tasks under removing obvious noise for further analysis.

1) *Muscle-finger system models*

We should tune the model that we select to simulate the real system and then analyze it through the trained model for modeling and analyzing the muscle-finger system. Implementing these concerns requires a white-box algorithm that can resolve variable relationships between dependent variables and multiple explanatory variables. As far as we know, the current generalized white-box algorithm is only multiple regression. Traditional multiple regression is usually converted into linear regression through linear processing. Then, the typical least square method is used to solve the preset parameters [11, 15, 16]. Thus, multiple regression is a liner regression essentially. In the EMG interface, there are too many EMG electrodes, which leads to too many items and high computational complexity [11, 15, 16]. Therefore, multiple regression generally uses first-order terms in EMG interface according to reference [17]. This paper compared Dendrite Net with first-order multiple regression, also known as multiple linear regression (LR). A 10-fold cross-validation (10-FCV) strategy was utilized for both LR and DD to evaluate the overall performance. It is worth pointing out that this paper focuses on the relation spectrum that "reads" the trained model of Dendrites Net. Thus, for poorly performing simplified models (LR), this paper just compared performance.

*a. Linear regression*

The linear regression is an algorithm that models the linear relationship [18]. For all subjects, we built the LR models bout single finger as following.

$$f(t) = LR(A^{lr}, E(t)) \quad (1)$$

Where $f(t)$ represents the finger force of single finger, $E(t)$ represents iEMG RMS of muscles, and $A^{lr}$ represents the regression coefficients.

*b. Dendrite Net*

Various NNs have been employed to model the relationship of the input space and the output space [19]. However, traditional NNs, called Cell body Net [11], are like black-box and provide a human-readable model.

Dendrite Net that imitates biological dendrites in brains is another novel basic machine learning algorithm [11]. Unlike machine learning algorithms that search for an appropriate classification curve or surface, Dendrite Net aims to design the logical extractor with controllable precision and is the white-box algorithm with lower computational complexity.

Here, we built a DD model of the muscle-finger system using three modules for each subject. The overall architecture of DD is shown in Figure 2. The architecture can be represented as the following formula.

$$\begin{cases} A^1(t) = W^{10} E(t) \circ E(t) \\ A^2(t) = W^{21} A^1(t) \circ E(t) \\ F(t) = W^{32} A^2(t) \end{cases} \quad (2)$$

Where $W^{10}$, $W^{21}$, and $W^{32}$ are the weight matrixes (Strength of synaptic connections), $A^1(t)$ and $A^2(t)$ are the output of DD modules.

$F(t) = [F_{ThumbF-E}(t) \; F_{ThumbA-A}(t) \; F_{Little}(t) \; F_{Ring}(t) \; F_{Middle}(t) \; F_{Index}(t)]^T$ represents the matrix of finger forces.

$E(t) = [1 \; E_{FDP}(t) \; E_{EDC}(t) \; E_{APL}(t) \; E_{FPL}(t) \; E_{EPL}(t) \; E_{EIP}(t)]^T$ represents the matrix of bias and iEMG RMS of muscles. $\circ$ denotes Hadamard product. The weight matrixes are solved by an error backpropagation.

## 2.3 Relation spectrum

*a. Fourier series* [20]

We start from trigonometric functions. Given the period $T=2l$, consider the harmonics

$$a_k \cos \frac{\pi k x}{l} + b_k \sin \frac{\pi k x}{l} \quad (1, 2, \cdots) \quad (3)$$

With frequencies $w_k = \frac{\pi k}{l}$ and periods $T_k = \frac{2\pi}{w_k} = \frac{2l}{k}$.

Since $T=2l=kT_k$, and an integral multiple of a period is again a period, the number $T=2l$ is simultaneously a period of all the harmonics. Thus, every sum of the form

$$s_n(x) = A + \sum_{k=1}^{n} \left( a_k \cos \frac{\pi k x}{l} + b_k \sin \frac{\pi k x}{l} \right) \quad (4)$$

Where $A$ is a constant, $\sum_{k=1}^{n} \left( a_k \cos \frac{\pi k x}{l} + b_k \sin \frac{\pi k x}{l} \right)$ is a function of period $2l$, since it is a sum of functions of period $2l$. The function $s_n(x)$ is called a trigonometric polynomial of order $n$ (and period $2l$). Then, the infinite trigonometric series can be expressed.

$$A + \sum_{k=1}^{\infty} \left( a_k \cos \frac{\pi k x}{l} + b_k \sin \frac{\pi k x}{l} \right) \quad (5)$$

Any function $f(x)$ can be decomposed into the sum of trigonometric functions.

$$f(x) = A + \sum_{k=1}^{\infty} \left( a_k \cos \frac{\pi k x}{l} + b_k \sin \frac{\pi k x}{l} \right) \quad (6)$$

These coefficients of trigonometric polynomial form Fourier spectrum. Later, the advent of the fast Fourier transform has greatly extended our ability to implement the Fourier spectrum on digital computers [21]. Today, it has become an essential tool to decompose signal.

*b. Taylor series* [22]

The Taylor series of a real or complex-valued function $f(x)$ infinitely differentiable at a real or complex number $a$ is the power series.

$$f(a) + \frac{f'(a)}{1!}(x-a) + \frac{f''(a)}{2!}(x-a)^2 + \frac{f'''(a)}{3!}(x-a)^3 + \cdots \quad (7)$$

In the more compact sigma notation, this can be written as

$$f(x) = \sum_{n=0}^{\infty} \frac{f^{(n)}(a)}{n!}(x-a)^n \quad (8)$$

Where $f^{(n)}(a)$ denotes the $n$-th derivative of $f$ evaluated at the point $a$.

Suppose we have gotten $m$ points of $f(x)$. The

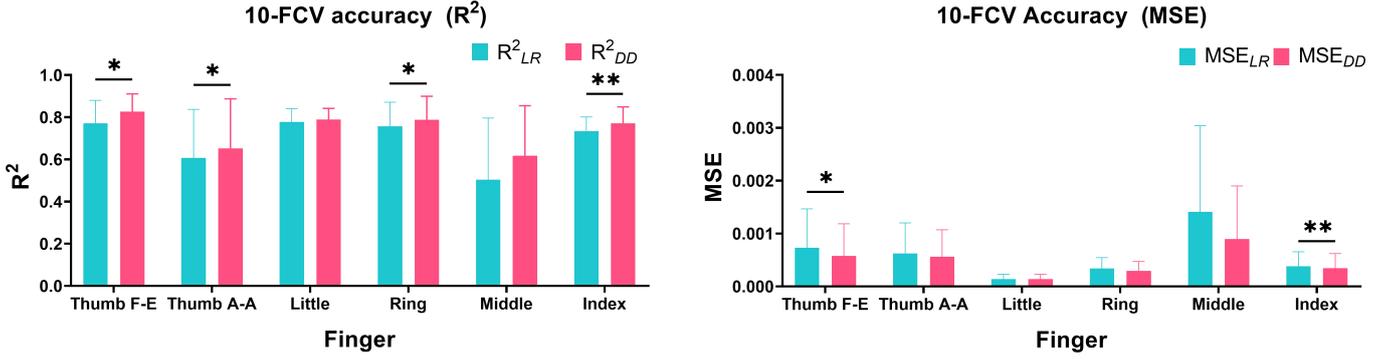

**Figure 3** 10-FCV accuracy of LR models and DD models. The coefficient of determination ($R^2$) and Mean-square error (MSE) were calculated to assess the models. P-values were calculated using paired samples t-tests. $^*P < 0.05$, $^{**}P < 0.01$. R2 or MSE: $mean \pm SD$.

Taylor expansion at each point is as follows.

$$\begin{cases} f(x) = \sum_{n=0}^{\infty} \frac{f^{(n)}(a_1)}{n!}(x-a_1)^n \\ \vdots \\ f(x) = \sum_{n=0}^{\infty} \frac{f^{(n)}(a_m)}{n!}(x-a_m)^n \end{cases} \quad (9)$$

Then, $f(x)$ can be expressed as including Taylor expansion with all sample points.

$$f(x) = \frac{1}{m}\left(\sum_{n=0}^{\infty} \frac{f^{(n)}(a_1)}{n!}(x-a_1)^n + \cdots + \sum_{n=0}^{\infty} \frac{f^{(n)}(a_m)}{n!}(x-a_m)^n\right)$$

$$= \sum_{n=0}^{\infty}\left(\frac{f^{(n)}(a_1)}{mn!}(x-a_1)^n + \cdots + \frac{f^{(n)}(a_m)}{mn!}(x-a_m)^n\right)$$

$$= \sum_{n=0}^{\infty}\left(\left[\frac{f^{(n)}(a_1)}{mn!} \cdots \frac{f^{(n)}(a_m)}{mn!}\right]\begin{bmatrix}(x-a_1)^n \\ \vdots \\ (x-a_m)^n\end{bmatrix}\right) \quad (10)$$

Eq. (10) can also be generalized to functions of more than one variable.

$$f(x_1,\cdots,x_d) = \sum_{n_1=0}^{\infty}\cdots\sum_{n_d=0}^{\infty}\left(\begin{bmatrix}\frac{\left[\frac{\partial^{n_1+\cdots+n_d}f}{\partial x_1^{n_1}\cdots\partial x_d^{n_d}}\right](a_{11},\cdots,a_{1d})}{m(n_1!\cdots n_d!)} \\ \vdots \\ \frac{\left[\frac{\partial^{n_1+\cdots+n_d}f}{\partial x_1^{n_1}\cdots\partial x_d^{n_d}}\right](a_{m1},\cdots,a_{md})}{m(n_1!\cdots n_d!)}\end{bmatrix}^T \begin{bmatrix}(x_1-a_{11})^{n_1}\cdots(x_d-a_{1d})^{n_d} \\ \vdots \\ (x_1-a_{m1})^{n_1}\cdots(x_d-a_{md})^{n_d}\end{bmatrix}\right) \quad (11)$$

Where $(x_d - a_{id})^{n_d}, i \in [1,\cdots m]$ can be simplified as the form that contains constant items and items containing $x_j, j \in [1,\cdots d]$. The expression in the generalized matrix form can be expressed as follows.

$$f(X) = T(W_{Taylor}, X) \quad (12)$$

Where $X = [x_1,\cdots,x_d]$, and $W_{Taylor}$ represents the coefficients matrix of polynomial. It is worth noting that $W_{Taylor}$ contains ***the derivatives at sample points in eq. (11)***. These are similar to those in Dendrite Net using backpropagation and chain rule. In essence, Dendrite Net also can be expressed as the generalized form.

$$f(X) = DD(W_{Dendritic\ Net}, X) \quad (13)$$

Where $W_{Dendritic\ Net}$ represents the weight matrix (Strength of synaptic connections). It is worth noting that $W_{Dendritic\ Net}$ contains ***the derivatives at sample points in eq. (11)***. $W_{Dendritic\ Net}$ of the trained DD can be translated into the relation spectrum about inputs and outputs by formula simplification with software (e.g., MATLAB or Python) [11]. The relation spectrum with a large number of coefficients can be expressed using a figure and a table, such as Figure 5 and Table 1. The position of items and coefficients are the abscissa and ordinate, and the items can be found in the table automatically generated by a computer. The relation spectrum expresses the impacts on outputs from the inputs, and the impacts contain independent and interaction effects in different degrees. It may become an essential tool to decompose a system or an online model. [*In fact, before this strategy, we had to use different models for offline analysis and online operation due to the black-box nature of online models, such as traditional neural networks model or support vector machines (SVM) model. Dendrite Net and relation spectrum can integrate offline analysis into the online operation or take an online model into the offline analysis.*] Additionally, orthogonal basis in signal decomposition of signal processing is to avoid repeated extraction of same power. The result of

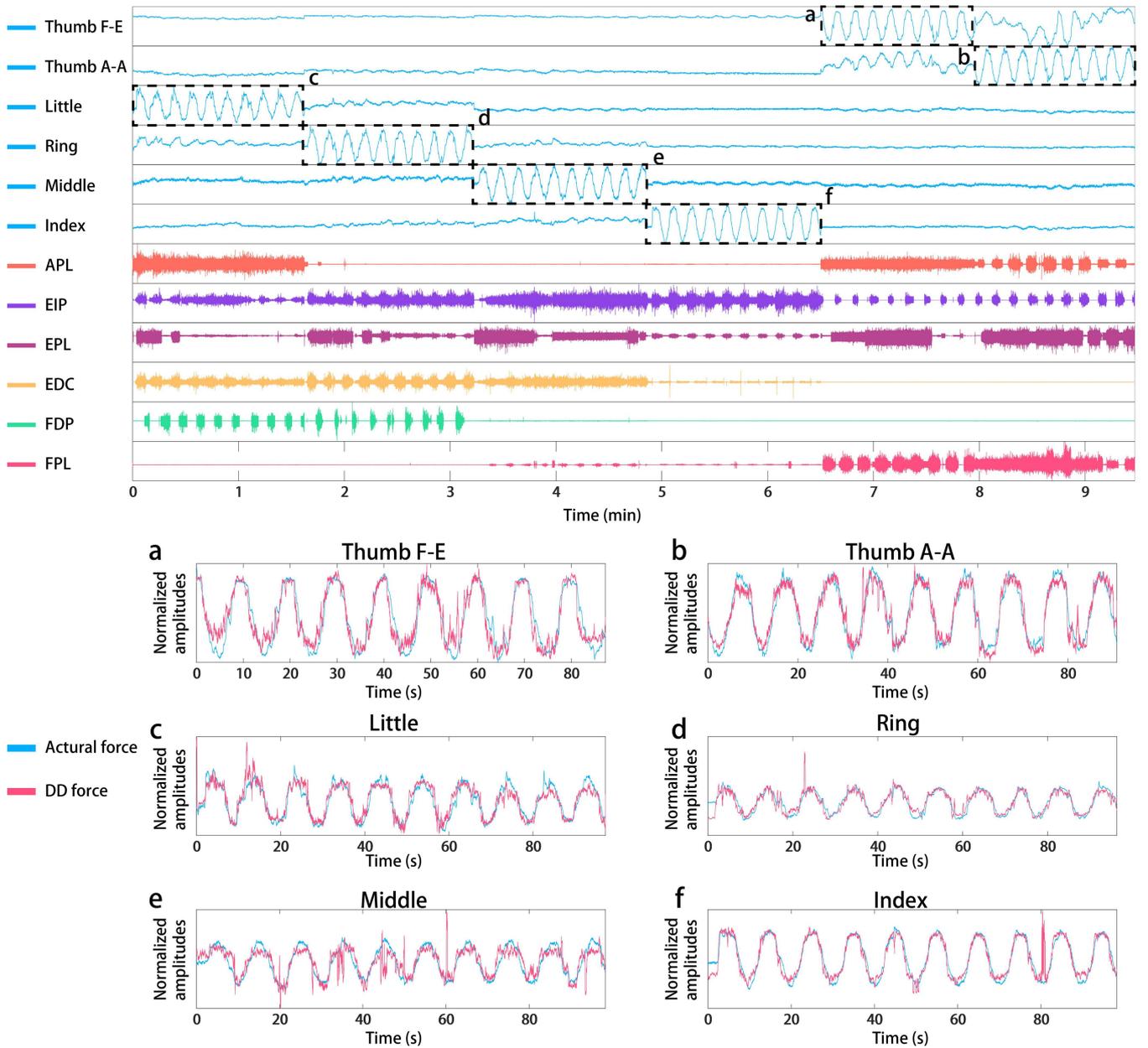

**Figure 4** Results of DD model for the muscle-finger system. As an example, the above figures showed results for testing data of Subject 3 in the 10-fold cross-validation.

Dendrite Net is showed as "addition". Therefore, no repeated extraction is present.

## 2.4 Identification of muscle synergies for single finger

We can solve the relation spectrum by formula simplification with software for the mechanism of muscle synergies on the fingers. This paper calculated each subject's relation spectrum that represents muscle synergies for a single finger and then found out the same contribution items. These same contribution items were useful to design the prosthetic hands with the intended functions that correspond to the physiologically appropriate muscles. The same contribution items $C(i)$ were defined as the following formula.

$$C(i) = \frac{Max\left(\left[\sum(co(i)>0) \quad \sum(co(i)<0)\right]\right)}{n} \times 100 \tag{14}$$

Where $n$ was the number of subjects, $co(i)$ was the coefficient of the corresponding item, $\sum(b)$ represented the number of meeting the conditions $b$.

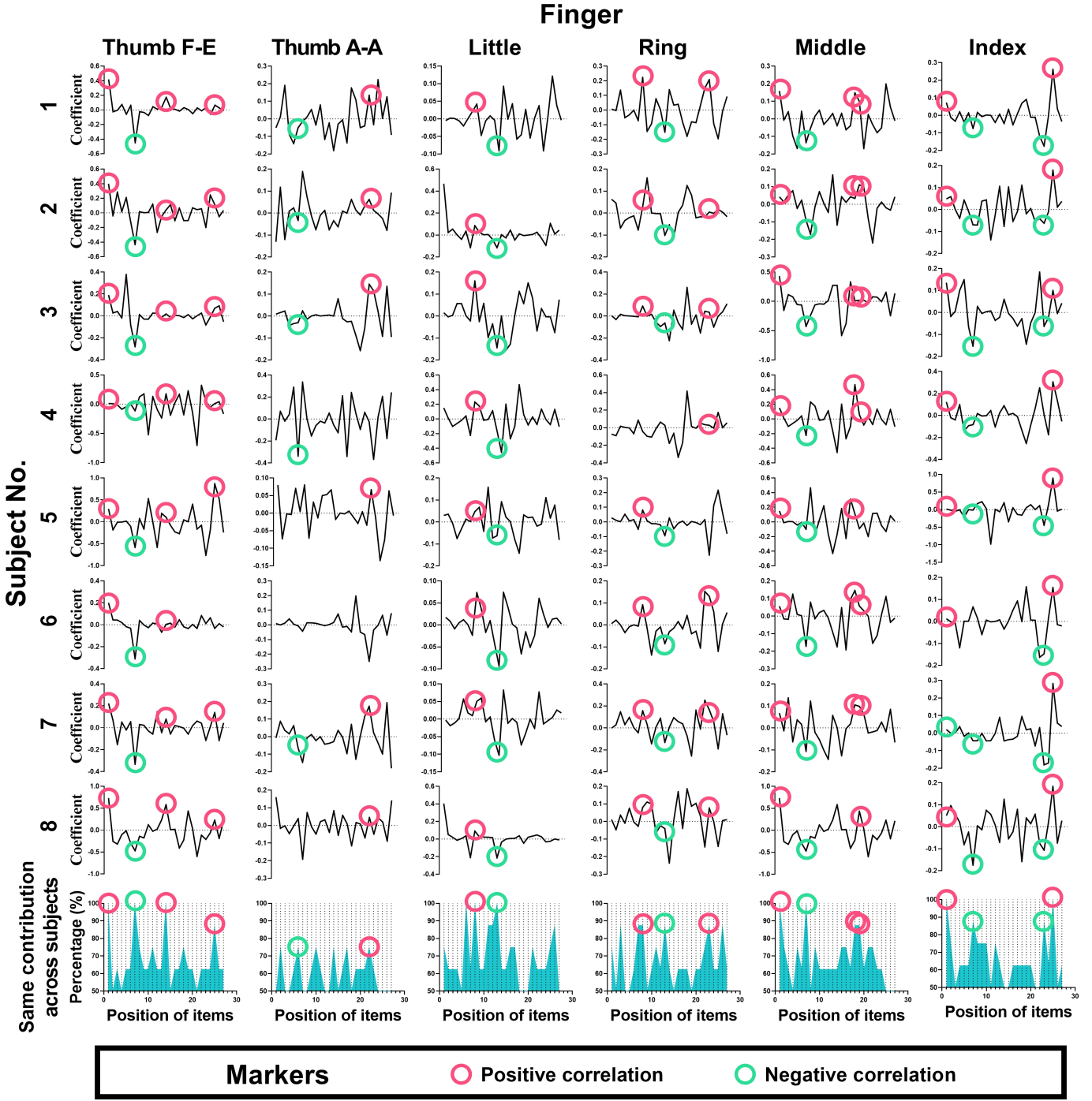

**Figure 5** Relation spectrum of the muscle-finger system in DD models. The definition of the same contribution items can be found in "Materials and Methods" Section [eq. (14)]. As an example, this paper marks some common items across subjects.

## 2.5 Identification of muscle coupling for hand

Because Dendrite Net has been transformed into the relation spectrum, we could analyze the muscle coupling for a hand, which has not been done before. Combining relation spectrums of all subjects, we calculated the correlation coefficient of the relation spectrum for each finger. These correlation coefficients meant muscle coupling of five-fingers and were calculated as the following formula [23].

$$R_{r^{f1}r^{f2}} = \frac{\sum_{i=1}^{n}\left(r^{f1}_{i} - \overline{r}^{f1}\right)\left(r^{f2}_{i} - \overline{r}^{f2}\right)}{\sqrt{\sum_{i=1}^{n}\left(r^{f1}_{i} - \overline{r}^{f1}\right)^{2}}\sqrt{\sum_{i=1}^{n}\left(r^{f2}_{i} - \overline{r}^{f2}\right)^{2}}} \quad (15)$$

**Table 1** Items in relation spectrum

| Position | Item | Position | Item |
|---|---|---|---|
| 1 | $E_{FPL}^2$ | 15 | $E_{EDC}E_{EPL}$ |
| 2 | $E_{FPL}E_{FDP}$ | 16 | $E_{EDC}E_{EIP}$ |
| 3 | $E_{FPL}E_{EDC}$ | 17 | $E_{EDC}E_{APL}$ |
| 4 | $E_{FPL}E_{EPL}$ | 18 | $E_{EDC}$ |
| 5 | $E_{FPL}E_{EIP}$ | 19 | $E_{EPL}^2$ |
| 6 | $E_{FPL}E_{APL}$ | 20 | $E_{EPL}E_{EIP}$ |
| 7 | $E_{FPL}$ | 21 | $E_{EPL}E_{APL}$ |
| 8 | $E_{FDP}^2$ | 22 | $E_{EPL}$ |
| 9 | $E_{FDP}E_{EDC}$ | 23 | $E_{EIP}^2$ |
| 10 | $E_{FDP}E_{EPL}$ | 24 | $E_{EIP}E_{APL}$ |
| 11 | $E_{FDP}E_{EIP}$ | 25 | $E_{EIP}$ |
| 12 | $E_{FDP}E_{APL}$ | 26 | $E_{APL}^2$ |
| 13 | $E_{FDP}$ | 27 | $E_{APL}$ |
| 14 | $E_{EDC}^2$ | 28 | 1 |

The 28 item is constant term, and it was not shown in Figure 5.

Where $R_{r^{f1}r^{f2}}$ represented the correlation coefficient between finger $f1$ and finger $f2$. $r^{f1}$ and $r^{f2}$ represented the relation spectrum of $f1$ and $f2$. $n$ was the number of items. An interesting result was shown in Figure 6.

## 3 Results

### 3.1 Muscle-finger system models

Figure 3 showed that the DD models outperformed the LR models in 10-FCV accuracy for the muscle-finger system, both in $R^2$ or MSE. As an example, the results of Subject 3 is shown in Figure 4. For thumb F-F, the data from Subject 2 and 4 were suspected outliers due to the electrode quality and were discarded [13]. [$R^2$: Thumb F-E: 0.771 ± 0.108 (LR) < 0.827 ± 0.084 (DD); Thumb A-A: 0.608 ± 0.230 (LR) < 0.653 ± 0.234 (DD); Little:0.777 ± 0.064 (LR) < 0.78 ± 0.053 (DD); Ring:0.757 ± 0.114 (LR) < 0.78 ± 0.111 (DD); Middle:0.505 ± 0.292 (LR) < 0.618 ± 0.237 (DD); Index:0.734 ± 0.067 (LR) < 0.770 ± 0.079 (DD).]

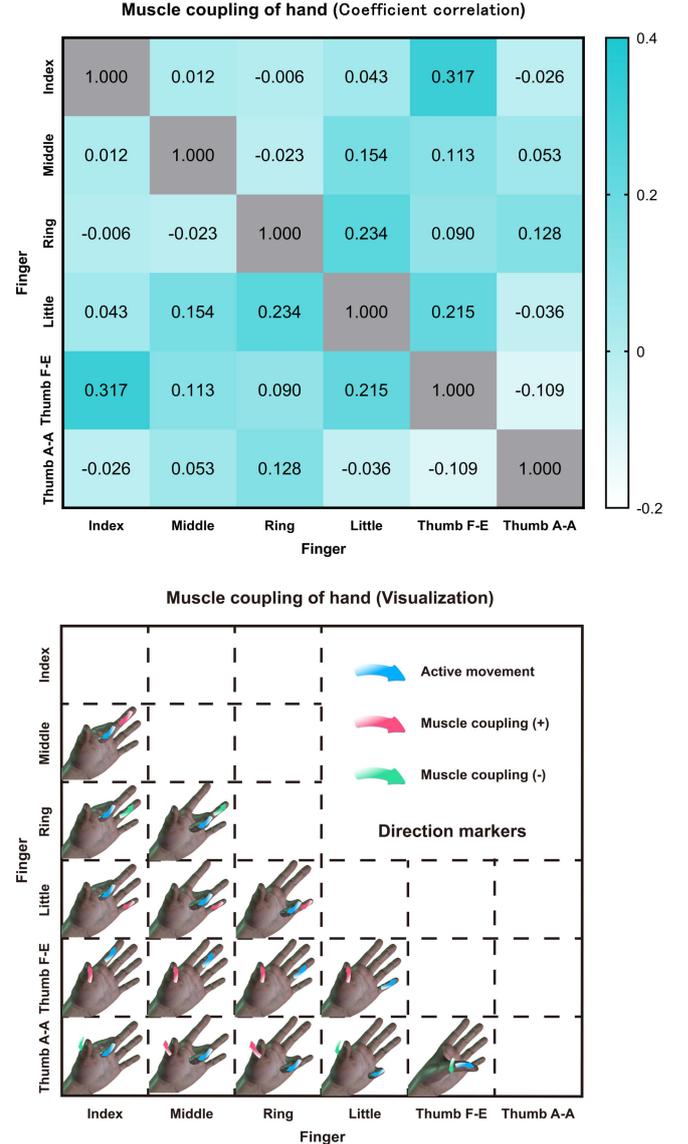

**Figure 6** Muscle coupling of hand. The results were interesting and could be verified by our own hands: 1. Raise our hand according to the perspective in the above figure. 2. Keep all fingers relaxed. 3. Move only the active finger and keep others relaxed. [Note that try our best to flex or extend to the maximum and feel it carefully. Due to the individual differences, the intensity of coupling may be different.]

### 3.2 Identification of muscle synergies for single finger

Since the advent of a neural network, it has been seen as a black box. In several decades, a neural network that can explain the relationship of input space and out space has been intensively sought.

For the muscle-finger system, the system has not been intuitive analysis due to its complex structure. This paper showed the relation spectrum of the muscle-finger system (see Figure 5). The relation spectrum can be read by the way of a checklist. "Position of items" corresponds to "Items" in Table 1. Despite some differences across subjects, a lot of the same contribution items existed in subjects.

### 3.2 Identification of muscle coupling for hand

The knowledge of muscle coupling for the hand is useful for the design of bionic prosthetic hands. Although the rough relationship between muscle and finger was found through anatomy and physiology, there was no method analyzing muscle coupling using online models prior to this study. This paper showed the muscle coupling of human hand. Despite differentiated coupling strength across subjects due to evolution, the common muscle coupling was shown in Figure 6. The results were interesting and could be verified by our own hands. [Note that we only move the active finger in Figure 6 and keep others relaxed when verifying the muscle coupling.]

## 4 Discussion

### 4.1 Specific engineering: muscle-finger system

As mentioned in the literature review, the knowledge of the intuitive link between muscle activity and the finger movement is conducive to the design of commercial prosthetic hands that do not need user pre-training. However, it is unclear about this link. The present study was designed to explore the intuitive link.

In terms of muscle synergies for a single finger, this study gave the relation spectrum about muscle activity and the finger movement. Some relations of this finding were consistent with the references [13, 14]. Nevertheless, our result was more precise. For example, previous research only showed that index finger movement was a positive correlation with the activation of EIP. However, this study not only revealed similar results to previous research but also showed that the index finger movement negatively correlated with the activation of FPL (see Figure 5 and Table 1). One unexpected finding was that the index finger movement was a negative correlation with the square of activation of EIP and positive correlation with the square of activation of FPL. This suggested that the intensity of the same muscle activation may affect the corresponding finger's movement direction. Besides, this phenomenon also showed that the relationship between muscle activation and finger movement was nonlinear, which explained why the DD models outperformed the simplified LR models. Because this paper focus on technology itself, the medical information would not be discussed too detailly. More details could be found from Figure 5 and Table 1. Additionally, because of the noise of electrode character and environment, the difference in signal strength, and each individual's specificity, there were some differences in the coefficient magnitude. Nevertheless, the relation spectrum is relatively similar across subjects.

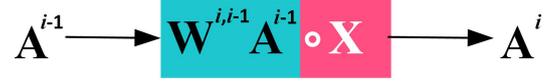

Figure 7 Generalized Dendrite module.

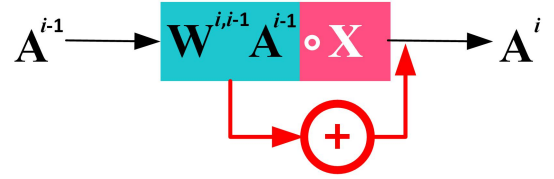

Figure 8 Residual Dendrite module.

Prior studies have noted the coupling phenomenon of hand [9, 10, 24]. The coupling of hand could be divided into the passive mechanical coupling and active muscle coupling. Passive mechanical coupling was measured by measuring implement [9] and has been used to design bionic hand [10]. This study was to assess active muscle coupling. These results about muscle coupling were in agreement with those obtained by an earlier study [9]. However, our results were more precise because our method was a quantitative analysis of iEMG DD models instead of using an analysis that subtracted the indices obtained in the passive condition from those obtained in the active condition [9]. More details could be found from Figure 6. These were particularly useful results for prosthetic hands.

### 4.2 Relation spectrum

In our previous study, we presented the generalized Dendrite Net [11]. The Dendrite module is expressed as follows (see Figure 7).

$$A^l = W^{l,l-1} A^{l-1} \circ X \qquad (16)$$

Where $A^{l-1}$ and $A^l$ are the inputs and outputs of the module. $X$ denotes the inputs of DD. $W^{l,l-1}$ is the weight matrix from the $(l-1)$-th module to the $l$-th module. $\circ$ denotes Hadamard product.

This paper demonstrated the similarity between the Taylor series and generalized Dendrite Net by proof-of-principle and presented the relation spectrum. The muscle-finger system's logical relationship was relatively simple, and we did not need more Dendrite modules. However, for a more complex model, we may use more dendrite modules. In order to solve the gradient vanishing problem, a residual strategy can be used [25]. The Residual

Dendrite module is expressed as follows (see Figure 8) [26].

$$A^l = W^{l,l-1}A^{l-1} \circ X + W^{l,l-1}A^{l-1} \qquad (17)$$

Similarly, Residual Dendrite Net (ResDD) also can be expressed in the generalized form essentially.

$$f(X) = DD(W_{Residual\ Dendritic\ Net}, X) \qquad (18)$$

Where $W_{Residual\ Dendritic\ Net}$ represents the weight matrix (Strength of synaptic connections). It is worth noting that $W_{Residual\ Dendritic\ Net}$ contains **the derivatives at sample points in eq. (11).** These derivatives in Residual Dendrite Net using backpropagation and chain rule are similar to those in Taylor series. Meanwhile, the calculation of Residual Dendrite Net only contains matrix multiplication, matrix addition, and Hadamard product. Thus, $W_{Residual\ Dendritic\ Net}$ of the trained ResDD can also be translated into the relation spectrum about inputs and outputs by formula simplification with software (e.g., MATLAB or Python) [11].

Additionally, traditional machine learning algorithms (e.g., NNs, SVM, or Decision tree) only generate a black-box model. Therefore, there are usually differences between the algorithms used in offline analysis and the online application, resulting in two phenomena. (1) Sometimes, the offline analysis shows promising results but poor performance online. (2) The online experiment shows poor performance, but we do not understand the reason at times. The relation spectrum's performance needs a large number of long-term applications to further verify in various fields in the future. However, we are pretty sure the relation spectrum shows the trained\online model using Dendrite Net. The online model can be "read" in offline analysis, which unifies online performance and offline results.

## 5 Conclusion

This paper demonstrated the similarity between the Taylor series and Dendrite Net using backpropagation and chain rule and then presented a relation spectrum. It is widely known that a functional relationship can be expressed by the sum of trigonometric or power polynomial. For the expression of a trigonometric polynomial, the typical example is Fourier frequency spectrum. Here, the relation spectrum is the spectrum of power series. The relation spectrum expresses the impacts on outputs from the inputs, and the impacts contain independent and interaction effects in different orders. Relation spectrum and Dendrite Net unify online performance and offline results.

In terms of specific engineering: we solved an unresolved but significant issue, the unclear link between muscle activity and finger movement, through Dendrite Net and relation spectrum [11]. The contribution lies in the relation spectrum of the muscle-finger system and the knowledge of muscle coupling, which may provide a reference for commercial prosthetic hands that do not need user pre-training.

Additionally, this paper was the first application of Dendrite Net, and we showed the concept of the relation spectrum from proof-of-principle for the first time systematically. Because Dendrite Net and Relation spectrum are both basic tools, they may be applied in most engineering fields in the future.